\documentclass{aastex}
\usepackage{spr-astr-addons}
\usepackage{url}
\usepackage{amsmath}
\usepackage{amsfonts}
\usepackage{amssymb}
\usepackage{graphicx}
\usepackage{subfigure}
\usepackage{graphicx}
\usepackage{dcolumn}
\usepackage{bm}
\RequirePackage{color}
\def\image1{\centerline{\color[gray]{.75}\rule{\hsize}{4pc}}}%

\begin{document}

\title{Chameleon mechanism with a new potential\\}
\slugcomment{Not to appear in Nonlearned J., 45.}
\shorttitle{Short article title} \shortauthors{Autors et al.}

\author{$^a$Kh. Saaidi\altaffilmark{1}} \author{$^b$H. Sheikhahmadi\altaffilmark{2}}\and \author{$^b$J. Afzali\altaffilmark{3}}
\affil{$^a$Islamic Azad University, Sanandaj branch,  Sanandaj, Iran.}
 \affil{$^b$Department of Physics, Faculty of Science, University of
Kurdistan,  Sanandaj, Iran}


\altaffiltext{1}{ksaaidi@uok.ac.ir.}
\altaffiltext{2}{h.sh.ahmadi@uok.ac.ir.}
\altaffiltext{2}{jamalafzali@gmail.com}
\begin{abstract}
 Chameleon scalar field is a new model, which introduced to provide a
mechanism for exhibiting accelerating universe. Chameleon field has
several interesting aspects, such as field dependence on the local
matter density. For this model we introduce a new kind of potential
which has run away form and satisfies chameleon constraints. The
results are acceptable in comparison with the other potentials which
studied up to now.
\end{abstract}



\section{Introduction}

The observation that universe appears to be accelerating at present
time has caused one of the greatest problem to modern cosmology. The
recent cosmological observation suggests that the universe consist
of about  $24\%$ cold dark matter and $76\% $ dark energy(DE)
\citep{1}, while DE has a negative pressure, is used to explain the
present cosmic acceleration. It is explicit that the nature of DE is
unknown for researchers until now, but they can describe it  by some
candidates. On of that candidates is Cosmological constant
$,\Lambda,$  but it has two well-known difficulties, the "fine
tuning" problem and the "cosmic coincidence" problem \citep{2}.
Observational data indicates that  $\Omega_{\Lambda}=0.763$ and
$\Omega_{m}=0.237$, so large value of $\Omega_{\Lambda}$ obviously
predicts that the universe is accelerating today, rather than
decelerating as had long been believed. The observation evidence
tells us that rate of expansion in the high-z region is slower than
that in our neighborhood. In this condition, where as variation of
the $\rho _{\Lambda }$  with respect to the time is equal to zero,
this is provide a problem in cosmology, called fine tanning, the
quintessence (cosmon-field) solved this problem, by using  coupling
between scalar field and dark matter \citep{3b,3}. There are several
different theories, which have been proposed by people, to interpret
the accelerating universe, such as, holographic DE model
\citep{4,41a,4b,41d}, agegraphic DE models \citep{5,a,b,c} and
scalar field models of DE, which including  quintessence field
\citep{61}, quintum field \citep{8a,8}, phantom field \citep{6} and
many others. While the quantity of cosmological constant is non
zero, the DE component is more generally modeled as quintessence
mechanism. It is a scalar field rolling down a flat potential. In
quintessence mechanism, the field has negative pressure and
therefore acts to accelerate expansion \citep{10}. Khoury and
Weltman (2003) have introduced another kind mechanism, which called
chameleon mechanism. In this mechanism the scalar field acquire a
mass whose magnitude depends on the local matter density \citep{11}.
Also it is a way to related an effective mass for scalar field
$\phi$. Scalar field is expansion field, and can be obtained from
string theory \citep{12,12a}. Also the chameleon mechanism is a way
to give an effective mass to a light scalar field via field self
interaction and interaction between field and matter \citep{13}. By
chameleon model they could detected fifth force that associated with
potential energy. We exhibit chameleon behavior for a new kind of
potential that  have not $\phi^4$ form at quintessence model
\citep{14,15}, but has a run away form. Where the consequences of a
run away potential for
chameleon mechanism can play the role of DE, we select our potential in this  category.\\
The scheme of the present paper is as follows: In sec. 2 we study
the preliminary of chameleon mechanism, by using two potentials, as
power law and exponential. In sec. 3 we introduce another potential,
where has run away form. For this model we obtain several parameters
which are useful in cosmology. One of that parameters is the matter
density in earliest time. The last section is devoted to conclusion.

\section{Preliminary}

We consider the general action as
\begin{eqnarray}\label{1}
S&=&\int
d^4x\sqrt{-g}\left(\frac{M_{pl}^{2}}{2}R-\frac{1}{2}(\partial\phi)
^{2}- V(\phi)\right)\cr  &-&\int
d^4x\sqrt{-g}\left(\frac{1}{\sqrt{-g}}\mathcal{L}_{m}(\psi_{m},g_{\mu\nu})\right)
\end{eqnarray}
where  $\phi $ is the chameleon scalar field and the potential
 $V(\phi)$ has run away form and $M_{pl}=(8\pi G)^{-\frac{1}{2}}$ = $2.44 \times10^{18}Gev$,
 is the reduced planck mass. Each matter field, $\psi$,  coupled to a metric in Jordan fame, is
related to the Einstein frame metric by a conformal transformation,
 $\widetilde{g}_{\mu\nu}=e^\frac{2\beta\phi}{M_{pl}}g_{\mu\nu}$. Here $ \beta$
 is the coupling constant without dimension. For the
matter that described by pressureless perfect fluid we have
$\tilde{g}^{\mu\nu}\tilde{T}_{\mu\nu}=-\tilde{\rho}$, where
\begin{eqnarray}\label{2}
\tilde{T}^{\mu\nu}= -\frac{2}{\sqrt{-\tilde{g}}}\frac{\partial
\mathcal{L}_{m}}{\partial \tilde{g}_{\mu\nu}},
\end{eqnarray}
and $\tilde{\rho}$ is the energy density. We need an effective
potential to govern the dynamic of the chameleon field. It can be
shown that\citep{18} :
\begin{equation}\label{3}
\nabla^{2}\phi =V_{eff,\phi}(\phi),
\end{equation}
where
\begin{equation}\label{4}
V_{eff}(\phi):=V(\phi)+\rho e^{\frac{\beta\phi}{M_{pl}}}.
\end{equation}
It is  necessary to be reminded that Khoury  and Weltman (2003 and
2004) had a good discussion  on chameleon model at two different
potentials, Ratra-Peebles and exponential  as
$$V(\phi)=\frac{M^{4+n}}{\phi^{n}},$$and$$V(\phi)=M^{4}exp(\frac{M^{n}}{\phi^{n}}),$$
respectively. Note that, for large value of the field, only power
law can tend to zero, see \citep{22,19,20} For further
 reviews. Also some researchers such as Brax (2004) and Waterhouse (2006)
 have considered chameleon model at different aspects for example
 DE, radion and chameleon  cosmology.
 We are  going to introduce another kind of potential
 which its results are acceptable and also satisfies chameleon condition.

\section{The Model}

We introduce a new potential such as

\begin{equation}\label{5}
 V(\phi)=\frac{a+b(q\phi)^{n}}{1+(q\phi)^{n}},
\end{equation}
where $a$ and $b$ are constants by dimension $Gev^{4}$,  q is a
 constant by dimension $Gev^{-1} $ and  $n$ is dimensionless real number.\\
This potential satisfies the constraints which emphasis in
\cite{14}, and then the asymptotic behavior of $V(\phi)$ is as
\begin{enumerate}
\item ${\lim_{\phi\rightarrow\infty} V(\phi)=b},$
\item ${\lim_{\phi\rightarrow0}V(\phi)=a},$
\item ${V_{,\phi}(\phi)}$ is  increasing  and  negative,
\item ${V_{,\phi\phi}(\phi)}$ is decreasing  and  positive.
\end{enumerate}
The most advantage of this potential is capability of bring
experiencing. Because one can obtain the constraints (1)...(4) by
different class of $a,b,q$ and $n$. This potential has quintessence
behavior because when $\phi$ tend to infinity it converge to $b$.
Here $b$ is very small and from this point of view, this potential
has prefer to exponential potential which is studied in
\citep{23,17}. The theoretical results got by this potential can be
very closed to observation evidence. Whereas this potential has
chameleon behavior, so it causes cosmic acceleration. This potential
 has a run away form, therefore effective potential has a minimum, so that from Eq. (\ref{3}), we have
\begin{equation}\label{6}
V_{eff,\phi}(\phi_{min})=0,
\end{equation}
therefore using (\ref{4}) and (\ref{6}), we have
\begin{eqnarray}\label{7}
\frac{nq^{n}\phi_{min}^{n-1}(b-a)}{(1+(q\phi_{min})^{n})^{2}}+
\rho\frac{\beta}{M_{pl}}e^{\frac{\beta\phi_{min}}{M_{pl}}}=0,
\end{eqnarray}
we can obtain the mass of the small fluctuations, $ m_{min}$ as
 \begin{eqnarray}\label{8}
 m_{min}^{2}=V_{,\phi\phi}(\phi_{min})
 +\rho\frac{\beta^{2}}{M_{pl^{2}}}e^{\frac{\beta\phi_{min}}{M_{pl}}},
 \end{eqnarray}
so that by substituting (\ref{6}) in (\ref{8}) we have
\begin{eqnarray}\label{9}
m_{min}^{2}&=&\frac{V_{,\phi}(\phi_{min})}{\phi_{min}}\left(n-1-\frac{2n(q\phi_{min})^{n}}
{1+(q\phi_{min})^{n}}\right)\cr
&+&\rho\frac{\beta^{2}}{M_{pl^{2}}}e^{\frac{\beta\phi_{min}}{M_{pl}}}.
\end{eqnarray}
Assuming that  universe is just  composed of dark matter and  DE, so
that by using the following data \citep{1,14}
$$\Omega_{matter}=0.237,~~~~~~~~~~~~~~~~~~\Omega_{DE}=1-\Omega_{matter}=0.763,$$
$$\rho_{matter}=1.04\times10^{-47}Gev^{4},~~~~~\rho_{DE}=3.34\times10^{-47}Gev^{4},$$
and $\rho=e^{\frac{3\beta\phi}{M_{pl}}}\widetilde{\rho}, $ in
conformal transformation, we can rewrite Eq. (\ref{7}) as
\begin{eqnarray}\label{10}
\frac{nq^{n}\phi_{min}^{n-1}(b-a)}{(1+(q\phi_{min})^{n})^{2}}+
\rho_{m}\frac{\beta}{M_{pl}}e^{\frac{4\beta\phi_{min}}{M_{pl}}}=0.
\end{eqnarray}
By getting the parameters of the Eq. (\ref{5}) as
$$a=1.1\times10^{-12}Gev^{4},~~~~~~~~~~~~~b=2.0\times10^{-48}Gev^{4},$$
$$q=2.05\times10^{20}Gev^{-1},~~~~~~~~~~~~n=0.9,~~~~~~\beta=9,$$
we obtain  $\phi_{min}=1.1775\times10^{18}Gev$. We  have drown
$V(\phi), V_{,\phi}(\phi)$ and $V_{,\phi\phi}(\phi)$, by these
constants for more introduction. From figure $(1)$ it is seen that
this potential satisfies the constraints of  \citep{23}. Our
definition, gives $V(\Phi)$ the dimensions of an energy density,
therefore, according to  \citep{10,14}, one can define DE density,
as $$\rho_{DE}=V(\phi_{min})=\frac{a+b(q\phi_{min})^{n}}{1+(q\phi_{min})^{n}},$$\\
by making use of $\phi_{min}$ and other constants $a, b, q $ and
$n$, we can obtain density of DE as
$\rho_{DE}=3.34\times10^{-47}Gev^{4}$. This result exactly is
equalled with main quantity  which is brought in \citep{14}. Now we
want obtain $m_{min}$ for this model. Assuming the matter is the
atmosphere of the earth with
$\widetilde{\rho}=4\times10^{-21}Gev^{4}$, using relation (\ref{9})
one can compute $m_{min}$ as
$$m_{min}=4.03\times10^{-24}Gev.$$
\begin{center}
\begin{figure}[ht]
\begin{minipage}[b]{1\textwidth}
\subfigure[\label{fig1a} ]{ \includegraphics[width=.25\textwidth]%
{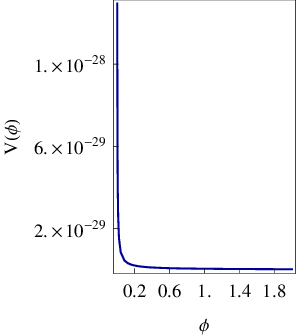}} 
\subfigure[\label{fig1b} ]{ \includegraphics[width=.25\textwidth]%
{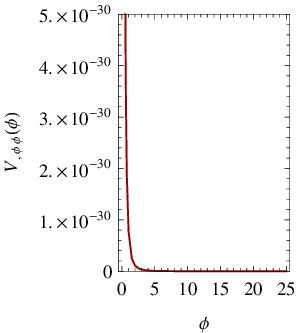}} \hspace{0cm}\\
\subfigure[\label{fig1c} ]{ \includegraphics[width=.25\textwidth]%
{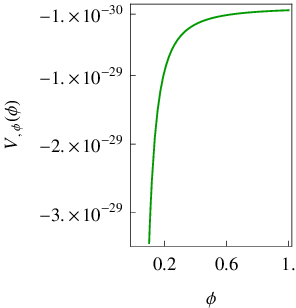}}
\end{minipage}
\caption{(a) is  $V(\phi)$, (b) is $V_{,\phi}(\phi)$ and (c) is
$V_{,\phi\phi}(\phi)$. In these figures we are use these constant,
$a=1.1\times10^{-12}Gev^{4}$, $b=2.0\times10^{-48}Gev^{4}$,
$q=2.05\times10^{20}Gev^{-1}$, $n=0.9$ and $\beta=9$.}
\end{figure}
\end{center}
We should note that $m_{min}$ is small fluctuation around minimum.
However, we obtain the  relation between $V_{,\phi}(\phi)$ and
momentum-energy tensor as
\begin{equation}\label{11}
T^{\mu\nu}=-\frac{2}{\sqrt{-{g}}}\frac{\partial\mathcal{L}_{m}}{{\partial
g_{\mu\nu}}}.
\end{equation}
In Jordan frame we have $\tilde{T}^{\mu\nu}
\tilde{g}_{\mu\nu}=3\tilde{p}-\tilde{\rho}$  , by substituting
$\tilde{p}=\omega\tilde{\rho}$ we have
$\tilde{T}^{\mu\nu}\tilde{g}_{\mu\nu}=(3\omega-1)\tilde{\rho}.$ But
in Einstein frame we obtain
\begin{equation}\label{12}
T^{\mu\nu}=\tilde{T}^{\mu\nu}e^{\frac{6\beta\phi}{M_{pl}}},
\end{equation}
from relation (\ref{12}) we have $ \tilde{T}^{00}\tilde{g}_{00}=
-\tilde{\rho}$, this means that
 $\omega=0$. Eventually we can obtain
\begin{equation}\label{13}
T^{00}=\tilde{\rho}e^{\frac{4\beta\phi}{M_{pl}}},
\end{equation}
by substituting (\ref{13}) in Eq.(\ref{10}), we have
\begin{equation}\label{14}
T^{00}=\frac{M_{pl}}{\beta}\left(\frac{n
q\phi_{min}^{n-1}(a-b)}{(1+(q\phi_{min})^{n})^{2}}\right),
\end{equation}
by making use of the value of $\phi_{min}$ and other relevant
constant we obtain $T^{00}=6.5\times10^{-48}Gev^{4}$ for this model.
This is another result which is agree with other works.\\
Now we want focus on the chameleon behavior in the earlier universe
by $\omega=-1$, of course note that in earlier universe, we use
\begin{equation}\label{15}
V_{eff}(\phi)=\frac{a+b(q\phi)^{n}}{1+(q\phi)^{n}}+\rho
e^{\frac{(1-3\omega)\beta\phi}{M_{pl}}},
\end{equation}
we define $\Omega_{m}$ as
$$\Omega_{m}=\frac{\rho_{m}}{\rho_{c}} e^\frac{\beta\phi_{min}}{M_{pl}},$$
where $\rho_{c}=3H^{2}M_{pl}^{2}$. Therefore from Eq. (\ref{9}) we
obtain
\begin{eqnarray}\label{16}
\frac{m_{min}^{2}}{H^{2}}&=&\frac{3\beta\Omega_{m}M_{pl}}{\phi_{min}}\left((1-n)
+\frac{2n(q\phi_{min})^{n}}{1+(q\phi_{min})^{n}}\right)\cr
&+&3\beta^{2}\Omega_{m},
\end{eqnarray}
for investigating  the cosmology behavior we consider two regimes,
\begin{itemize}
\item   $\phi_{min}\geq b^{\frac{1}{4}}$\\
\item   $\phi_{min}\gg b^{\frac{1}{4}}$
\end{itemize}
For $\phi_{min}\geq b^{\frac{1}{4}}$  regime, where
$(q\phi_{min})^{n}\simeq(10^{17})^{n}$ is very larger than one in
denominator, and $\frac{M_{pl}}{\phi_{min}} \simeq 10^{39}$, so that
we can rewrite Eq.(\ref{16}) as
\begin{eqnarray}\label{17}
\frac{m_{min}^{2}}{H^{2}}\simeq 3\beta\Omega_{m}(n+1)\times10^{39},
\end{eqnarray}
it is well known for $\Omega_{m}> 10^{-28}$ we have
$\frac{m_{min}^{2}}{H^{2}}\gg1.$ We can obtain the similar result
for the case which we have coupling constant cosmology  only. In
this case,  equation of state is $P=-\rho$ this means that
$\omega=-1$, therefore according to Eq.({\ref{9}) one can obtain the
similar result by replacing $4\beta$ instead of $\beta$. So that we
have
\begin{eqnarray}\label{18}
\frac{m_{min}^{2}}{H^{2}}&=&
\frac{12\beta\Omega_{vac}M_{pl}}{\phi_{min}}\left((1-n)+
\frac{2n(q\phi_{min})^{n}}{1+(q\phi_{min})^{n}}\right)\cr
&+&48\beta^{2}\Omega_{vac},
\end{eqnarray}
therefore we can obtain $\frac{m_{min}^{2}}{H^{2}}\simeq48
\beta^{2}\Omega_{vac}\gg1$, as
$$\Omega_{vac}=\frac{\rho_{vac}e^{\frac{4\beta\phi_{min}}{M_{pl}}}}{\rho_{c}}.$$
For $\phi_{min}\gg b^{\frac{1}{4}}$  regime, from  Eq.(\ref{16}), we
have
\begin{eqnarray}\label{19}
\frac{m_{min}^{2}}{H^{2}}\simeq\frac{3\beta\Omega_{m}M_{pl}}{\phi_{min}}(n+1),
\end{eqnarray}
By using the  definitions  of $\Omega_{m}$, we obtain
\begin{equation}\label{20}
\frac{m_{min}^{2}}{H^{2}}
\simeq\frac{3{\beta}\rho_{m}e^{\frac{\beta\phi_{min}}{M_{pl}}}}{\rho_{c}\phi_{min}}(n+1),
\end{equation}
using $\rho_{c}=3H_{0}^{2}M_{pl}^{2}$, and Eq.(\ref{20}), we have
\begin{equation}\label{21}
\rho_{m}=\frac{m_{min}^{2}\phi_{min}M_{pl}}{(n+1)e^{\frac{\beta\phi_{min}}{M_{pl}}}},
\end{equation}
By making use of the value of $m_{min}$ and other introduced
constant, we can arrive at
$$\rho_{m}=0.4\times10^{-16} Gev^{4}.$$ \\  This result is an estimate for matter density in earliest universe.
Note that in present time $\rho_{m}\simeq 10^{-47}Gev^{4}$ therefore
our estimate  says that the density falls off in proportion to the
volume of the universe. Also  this condition
 is for earliest time, and  $\phi_{min}$ increase with time, so that the matter density is diluted.
We see that by this potential with out complex computation the
results  be satisfied.

\section{Conclusion}
In this paper,  we  introduced a potential which satisfy the
chameleon mechanism conditions. We obtain a situation that this
potential is become run away form. It is notable that this kind of
potential has quintessence behavior, because when $\phi$ tend to
infinity it converge to $b$ which is very small and nearly equal to
zero. This properties prefer our model to exponential type. By
making use of this kind of potential, we have  obtained  the DE
density, minimum mass of scalar field and $\phi_{min}$ as $\rho_{DE}
= 3.34 \times 10^{-47} Gev^4$, $m_{min}= 4 \times 10^{-21}Gev^4$ and
$\phi_{min} = 1.1775 \times 10^{18}Gev$ respectively.  It is seen
that these obtained values are in comparison with observational
data. Also, we have investigated two regimes of $\rho_m$ on the
present and earliest time. We have found $\rho_m = 0.4 \times
10^{-16} Gev^4$ for matter density in early time. Our results are in
accordance with other articles \citep{17,18,14}.

\end{document}